\documentclass[fleqn,usenatbib]{mnras}

\usepackage{newtxtext,newtxmath}

\usepackage[T1]{fontenc}

\DeclareRobustCommand{\VAN}[3]{#2}
\let\VANthebibliography\thebibliography
\def\thebibliography{\DeclareRobustCommand{\VAN}[3]{##3}\VANthebibliography}

\usepackage{graphicx, float, hyperref, tabularx, physics}	
\usepackage{amsmath}	
\usepackage{amssymb}	

\usepackage{threeparttable} 

\newcommand{\grs}{{GRS~1915+105 }}
\newcommand{\rxte}{{\it{RXTE} }}

\title[High-frequency variability in GRS~1915+105]{The evolution of the high-frequency variability in the black hole candidate GRS~1915+105 as seen by \textit{RXTE}}

\author[Y.\ Zhang et al.]{
Yuexin Zhang,$^{1}$\thanks{E-mail: yzhang@astro.rug.nl}
Mariano M\'{e}ndez,$^{1}$
Federico Garc\'{i}a,$^{1,6}$
Konstantinos Karpouzas,$^{1,2}$
Liang Zhang,$^{3,2}$
\newauthor
Honghui Liu,$^{4}$
Tomaso M.\ Belloni,$^{5}$
and Diego Altamirano$^{2}$
\\
$^{1}$Kapteyn Astronomical Institute, University of Groningen, P.O. BOX 800, 9700 AV Groningen, The Netherlands\\
$^{2}$School of Physics and Astronomy, University of Southampton, Southampton, SO17 1BJ, UK\\
$^{3}$Key Laboratory of Particle Astrophysics, Institute of High Energy Physics, Chinese Academy of Sciences, Beijing 100049, People's Republic of China\\
$^{4}$Center for Field Theory and Particle Physics and Department of Physics, Fudan University, 200438 Shanghai, China\\
$^{5}$INAF - Osservatorio Astronomico di Brera, via E. Bianchi 46, I-23807 Merate, Italy\\
$^{6}$Instituto Argentino de Radioastronom\'ia (CCT La Plata, CONICET; CICPBA; UNLP), C.C.5, (1894) Villa Elisa, Buenos Aires, Argentina\\
}

\date{Accepted XXX. Received YYY; in original form ZZZ}

\begin{document}
\label{firstpage}
\pagerange{\pageref{firstpage}--\pageref{lastpage}}
\maketitle

\begin{abstract}
GRS~1915+105 can show type-C quasi-periodic oscillations (QPOs) in the power density spectrum. A high-frequency QPO (HFQPO) at 67~Hz has been observed in this source, albeit less often than the type-C QPOs. Besides these features, GRS~1915+105 sometimes shows a broad bump in the power spectrum at around 30--150~Hz. We study the power spectra of GRS~1915+105 with the \textit{Rossi X-ray Timing Explorer} when the source was in the $\chi$ class. We find that the rms amplitude of the bump depends strongly upon both the frequency of the type-C QPO and the hardness ratio, and is correlated with the corona temperature and anti-correlated with the radio flux at 15~GHz. The characteristic frequency of the bump is better correlated with a combination of the frequency of the type-C QPO and the hardness ratio than with the frequency of the type-C QPO alone. The rms amplitude of the bump generally increases with energy from $\sim$1--2\% at $\sim$3~keV to $\sim$10--15\% at $\sim$30~keV. We suggest that the bump and the high-frequency QPO may be the same variability component but the properties of the corona affect the coherence of this variability, leading either to a HFQPO when the spectrum is in the relatively soft $\gamma$ class, or to a bump when the spectrum is in the hard $\chi$ class. Finally, we discuss the anti-correlation between the rms amplitude of the bump and the radio flux in the context of the relation between the corona and the jet.
\end{abstract}




\section{Introduction}

Black hole binaries (BHBs) can show variability in the X-ray light curve on timescales from milliseconds to years~\citep[e.g.,][]{2014SSRv..183...43B,2016ASSL..440...61B,2019NewAR..8501524I}. Fourier techniques are a powerful tool to explore the variability of X-ray time series~\citep{1989ASIC..262...27V}. Power density spectra (PDS) are generally used to study the variability of the light curve in the frequency domain, where several components can be observed. Some of those components are the broadband noise, that can extend up to 10--100~Hz, and narrow peaks called quasi-periodic oscillations~\citep[QPOs;][]{2000MNRAS.318..361N,2002ApJ...572..392B,2005AN....326..798V,2020arXiv200108758I,2021ASSL..461..263M}. The low-frequency QPOs (LFQPOs; $\sim$mHz to $\sim$30~Hz) in black-hole X-ray binaries can be divided into three classes, i.e.\ type A, B, and C, based on the quality factor, fractional root mean square (rms) amplitude, phase lags of the QPOs and the strength of the underlying broadband noise component~\citep{2005ApJ...629..403C}. High-frequency QPOs in black holes X-ray binaries with frequencies up to $\sim$350~Hz are less common and were only observed in a few sources~\citep[e.g.,][]{1997ApJ...482..993M,1999ApJ...522..397R,2001ApJ...552L..49S,2012MNRAS.426.1701B,2013MNRAS.435.2132M}. High-frequency QPOs and high-frequency broadband component with a cutoff at 60--80~Hz~\citep{2001ApJ...558..276T} in black-hole X-ray binaries and kilo-hertz QPOs in neutron-star X-ray binaries have many similarities, indicating that they may have the same origin~\citep{1999ApJ...520..262P,2002ApJ...572..392B,2013MNRAS.435.2132M,2021MNRAS.508.3104B}.

The rms amplitude of the components in the PDS can be used to study the correlated X-ray variability in these sources. For instance, the 0.1--10~Hz integrated rms amplitude in the power spectrum can be regarded as a tracer of accretion regimes in black hole transients~\citep{2011MNRAS.410..679M}. The rms amplitude of the QPOs increases with energy, gradually flattens above 10~keV, and at the highest energies it can either increase or decrease~\citep[e.g.,][]{2013MNRAS.434...59Y,2017ApJ...845..143Z,2020MNRAS.494.1375Z,2020ApJ...895..120L}. Phase lags are defined as the argument of the Fourier cross spectrum of two light curves at different energies over a fixed frequency range, and can be used to study other properties of the signals that we identify in the power density spectrum~\citep{1999ApJ...510..874N}. The lags of QPOs can increase, decrease, or remain constant with energy~\citep[e.g.,][]{2000ApJ...541..883R,2017ApJ...845..143Z,2020MNRAS.494.1375Z}.

Low mass X-ray binaries (LMXBs) emit in the X-ray band as material accretes onto the compact object. A typical X-ray spectrum includes a power-law component, a disk thermal component, and a reflection component, with the strength of the different components depending on the state of the source~\citep[for a review, see][]{2006ARA&A..44...49R}. The thermal disk is represented by a multi-temperature blackbody with a temperature around 0.3--2.0~keV. When the soft photons interact with electrons in the hot corona near the black hole, inverse Compton scattering transfers energy from the hot corona to the thermal photons and gives rise to the power-law emission with a high-energy cutoff. A fraction of the Comptonized photons illuminate the disk and are Compton back-scattered, producing fluorescent emission lines and a Compton hump. In the hard state, Comptonization is strong, yielding a power law that is flat and can extend to high energies, and reflection usually appears, while in the soft state, the power law is steeper and a stronger disk thermal spectrum is present.

GRS~1915+105 is a black-hole X-ray binary discovered in 1992 by WATCH~\citep{1992IAUC.5590....2C}. As the first galactic micro-quasar, GRS~1915+105 shows strong variability in the X-ray band and a luminous jet in the radio band~\citep[for a review, see][]{2004ARA&A..42..317F}. Typically the outburst of a black-hole LMXB undergoes an evolution along a `q' path on the hardness intensity diagram~\citep[HID;][]{2004MNRAS.355.1105F}: The outburst starts with the source in the quiescent state, continues when the source enters the low hard state, the source quickly transits to the high soft state, and finally returns to the quiescent state; a relativistic jet appears between the hard and the soft states. GRS~1915+105 is peculiar since it does not show this typical behavior of black hole transients, and never goes back into quiescence since it was discovered 30 years ago, even though in the last two years this source displayed an extreme low flux state and indication of rebrightening~\citep{2018ATel11828....1N,2020ATel13652....1A,2020ATel13676....1L}. \citet{2020ApJ...904...30M} suggested that this low flux state of GRS~1915+105 is not quiescence but a highly absorbed, quasi compton-thick accretion mode, and \citet{2021MNRAS.tmp..538M} suggested that this low flux epoch might be associated to super-Eddington accretion.

Due to its variability and persistence, tremendous work has been done on GRS~1915+105. \citet{2000A&A...355..271B} performed a model-independent analysis and classified the variability into twelve classes, with two more classes added by~\citet{2003A&A...411L.415H,2005A&A...435..995H}. The low-frequency QPOs in GRS~1915+105 have been intensively studied in the past~\citep[e.g.,][]{2000ApJ...541..883R,2013MNRAS.434...59Y,2015MNRAS.446.3516I,2020MNRAS.494.1375Z,2020arXiv201201825L}. The rms amplitude of the LFQPOs increases with energy and then gradually flattens, indicating that the energy that sustains the QPOs possibly originates from the hot corona~\citep{2013MNRAS.434...59Y}. The slope of the lag-energy spectrum of the QPOs changes monotonically from positive to negative as the QPO frequency increases, with the slope becoming 0 at around 2~Hz, where the phase lags are more or less independent of energy. At the same time the rms amplitude of the QPOs increases and then decreases as the QPO frequency increases, with a maximum at 2~Hz~\citep{2000ApJ...541..883R,2010ApJ...710..836Q,2013ApJ...778..136P,2020MNRAS.494.1375Z}. HFQPOs (either around 35~Hz or 70~Hz) in GRS~1915+105 are less frequent, and appear depending on the spectral states and the variability class of the source~\citep{2006MNRAS.369..305B,2013MNRAS.432...10B,2013MNRAS.435.2132M}.

\cite{2001ApJ...558..276T} used data from the \textit{Rossi X-ray Timing Explorer}~\citep[\textit{RXTE};][]{1993A&AS...97..355B} from 1996--2000 to explore a high-frequency feature in the power density spectra of GRS~1915+105. This author identified two types of steady hard X-ray states of GRS~1915+105, which are associated to the ``plateau'' state and the $\chi$ class~\citep{1999MNRAS.304..865F,2000A&A...355..271B}. In the type I state, the energy spectrum shows a power-law continuum with a cut-off energy between 60 and 120~keV. The power density spectrum shows a significant high-frequency noise component, which \citet{2001ApJ...558..276T} fitted with a Gaussian function, with a width of around 60--80~Hz. In the type II state, the power-law component that fits the energy spectrum shows a break at an energy between 12 and 20~keV. The high-frequency part of the power density spectrum drops quickly, showing no significant variability at frequencies higher than $\sim$30~Hz. These two types of X-ray hard states can also be distinguished  by their properties in the radio band. In the type I state, the source tends to be ``radio quiet'', while in the type II state, the source shows high radio flux. \citet{2001ApJ...558..276T} also gave a quantitative estimate of the properties of the high frequency bump, assuming that the QPO frequency and the width of the high-frequency bump are proportional to the Keplerian frequency at the outer and inner radii of an optically thin region of the accretion flow. This estimate matches the fitting results of the power density spectra.  We note that, besides the observations between 1996--2000, \rxte collected a large amount of data on GRS~1915+105 before the termination of the mission in 2012. \citet{2022NatAs.tmp...51M} recently analyzed these same data and suggested that the X-ray corona turns into the jet.

In this paper, we explore the full \rxte archival data of GRS~1915+105 in detail to further study the high-frequency bump in the power spectrum, complementing and expanding the work of~\citet{2022NatAs.tmp...51M}. The paper is organized as follows: In Section~\ref{sec:data} we describe the reduction of the \rxte data of GRS~1915+105, and explain how we generated the power density spectra as well as the cross spectra for different energy bands. We also explain how we filter and analyze the data in this section. In Section~\ref{sec:results} we show the results we obtain. These results include the power density spectra, the rms evolution of the high-frequency bump as a function of the QPO frequency and hardness ratio, and the rms energy evolution for the combined fitting. In Section~\ref{sec:discussion} we discuss our results and compare our results with previous work. In Section~\ref{sec:conclusion} we summarise our conclusions.

\section{Observations and Data Analysis}\label{sec:data}

\rxte monitored \grs from 1996 to 2012. There are more than 1800 public observations of \grs in the \rxte archive.

Among all the \rxte observations of GRS~1915+105, we analyzed the 620 in the $\chi$ state~\citep{2000A&A...355..271B} with type-C QPOs, following the same selection as in~\citet{2020MNRAS.494.1375Z} and \citet{2022NatAs.tmp...51M}. Finally we chose 410 observations which have ``simultaneous'' (within 2 days) radio observations. For our analysis we used the Event, Single Bit and Binned mode files, with a time resolution of $1/512$~s so that the Nyquist frequency is 256~Hz. We used \textit{Ghats}~\footnote{\url{http://www.brera.inaf.it/utenti/belloni/GHATS_Package/Home.html}} to generate the power density spectrum (PDS) for the full Proportional Counter Array~\citep[PCA;][]{1993A&AS...97..355B} band, channels 0--249; in some observations the measurements start at channel 8, so the full PCA band in those cases is channels 8--249. We set the length of each Fast Fourier Transformation (FFT) segment to 16~s. We averaged the PDS of all time segments within one observation, subtracted the Poisson level~\citep{1995ApJ...449..930Z}, and normalized the PDS to units of rms$^{2}$ per Hz~\citep{1990A&A...230..103B}. The background count rate can be ignored for the full PCA channels since it is negligible compared to the source count rate. Finally, we applied a logarithmic rebin in frequency to the PDS data such that the size of a bin increases by $\exp(1/100)$ compared to that of the previous one.

We separated the full PCA bands into 5 bands defined by the channels: 0--13, 14--35, 36--$\sim$50, $\sim$51--$\sim$103, $\sim$104--249, where the ``$\sim$'' indicates the nearest channel available. For some observations with Single Bit Mode, the data started from channel 8, so the first energy band spanned channels 8 to 13. The \textit{RXTE}/PCA energy-channel conversion is divided into 5 epochs in time sequence~\footnote{\url{https://heasarc.gsfc.nasa.gov/docs/xte/e-c_table.html}}. The observations we analyzed in this paper are from epochs 3, 4, and 5. We can obtain the energy boundaries of each channel interval by mapping the channels based on the time of each observation. Different from the full band, in the subsets of channels, we needed to take the background rate into account to normalize the PDS to rms units. We used \texttt{pcabackest} to calculate the background count rate per Proportional Counter Unit (PCU) for each channel range for the three epochs, as well as the number of PCU that was active in each observation.

For each observation, we also computed the FFTs for different energy bands in order to obtain the cross spectra, estimate the coherence function, and calculate the phase lags that is energy- and frequency-dependent~\citep{1997ApJ...474L..43V,1999ApJ...510..874N}. We regarded the lowest energy band (channels 0 to 13 or 8 to 13) as the reference band. In this process, we used the same files with 1/512-s time resolution and 16-s time segment as we did for the power density spectra. We calculated the intrinsic coherence between the FFTs as a function of frequency before we calculated the phase lags.

After subtracting the background, we obtained the observed count rate in the 13--60~keV (hard) band and the 2--7~keV (soft) band, respectively, using the closest channels that match these energy ranges, in units of the Crab nebula as in~\citet{2008ApJ...685..436A}. We divided the hard band intensity by the soft band intensity to calculate the hardness ratio.

The radio data of GRS~1915+105 were obtained with the Ryle Telescope at 15 GHz~\citep{1997MNRAS.292..925P}. Calibrations were performed on the flux density with the help of observations of the quasars 3C~48 or 3C~286. The observations contain mostly 32-s samples with an rms noise of 6~mJy, which decreases as the square root function of the integration time, while the flux density below about 1~mJy is possibly unreliable. See~\citet{1997MNRAS.292..925P} for more discussion about the radio data.

\subsection{Fitting of the power density spectra}\label{subsec:fit}

We used XSPEC version 12.9~\citep{1996ASPC..101...17A} to fit the power spectra. PYTHON scripts were used to assist performing the semiautomatic fitting of the 410 observations. A number of Lorentzian functions and a Gaussian function were used to fit the variability of the source~\citep{2000MNRAS.318..361N, 2002ApJ...572..392B, 2001ApJ...558..276T, 2020MNRAS.494.1375Z}. Four Lorentzian functions were used to represent the fundamental, harmonic, and sub-harmonic of the type-C QPOs plus the broad-band noise. We used two extra Lorentzian functions to fit the power spectra that were not satisfactorily fitted by these four Lorentzian functions. For instance, some power spectra showed two broadband components, one at low and another at high frequency, which we fitted with zero-centered Lorentzians. In some power spectra the QPOs themselves moved to some extent during an observation; in those cases one Lorentzian function was not enough to fit the QPOs. We noticed that power spectra with the type-C QPOs at the same frequency sometimes showed enhanced power at frequencies between $\sim$30~Hz and $\sim$150~Hz. Following~\citet{2001ApJ...558..276T} we added a Gaussian centered at zero to fit this excess power. We left the width of the Gaussian and the integrated power of the Gaussian free during the fits. We name this component the high-frequency bump, or simply the bump.

We converted the normalization of all components into fractional rms amplitude, hereafter rms~\citep{1990A&A...230..103B}:
\begin{equation}
    \text{rms} = \frac{\sqrt{P(S+B)}}{S}, \label{eq:rms}
\end{equation}
where $P$ is the normalization of the Lorentzian giving the integrated power of the bump in Leahy units~\citep{1983ApJ...266..160L}, and $S$ and $B$ are, respectively, the source and background count rates. For the error of the rms, we considered also the uncertainties of the source and background count rates. Unless indicated otherwise, all errors represent the 90\% confidence level of the corresponding quantity.
Because some components were not significant, in some cases we calculated the 95\% upper limit. We estimated the significance of the bump as its integrated power divided by its error bar of 1-$\sigma$ (68\%) confidence level. If this factor was lower than 3, we considered that the bump was not significant. For those cases, we fixed the width of the Gaussian function at 70~Hz and gave the 95\% upper limit of the rms. Finally, when the bump was significant but out of the 30--150~Hz range, we fixed the width of the Gaussian at 70~Hz and computed the upper limit of the rms. We carried out this same process in each energy band (see above). We noticed that the PDS for channels from $\sim$104 to 249 showed no significant signal, therefore we did not consider this band in our analysis. We used the best-fitting model for each observation as a baseline to perform further fitting on the four separate bands.



\section{Results}\label{sec:results}

\begin{figure*}
    \includegraphics[width=0.49\textwidth ,trim= 30 30 30 30]{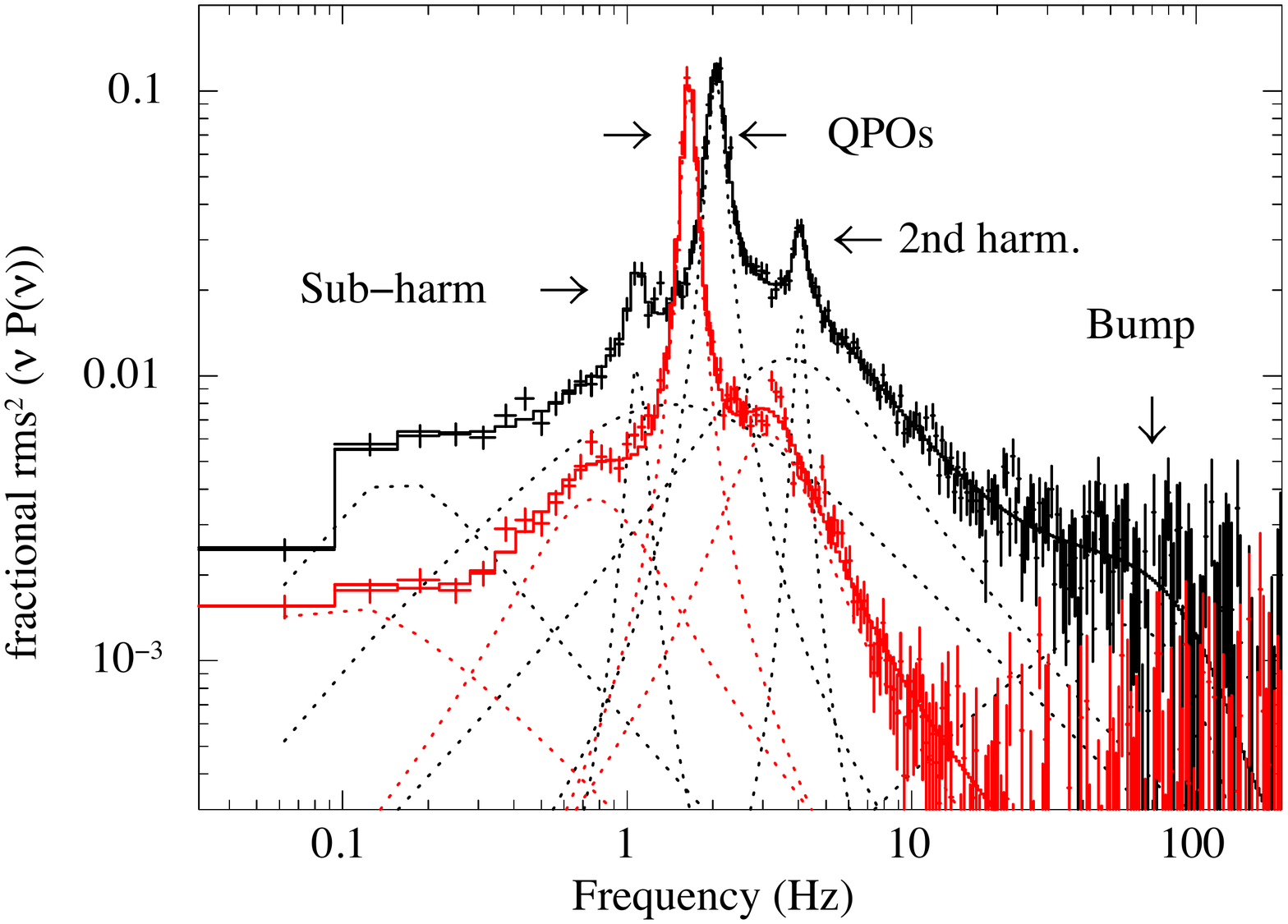}
    \includegraphics[width=0.49\textwidth ,trim= 30 30 30 30]{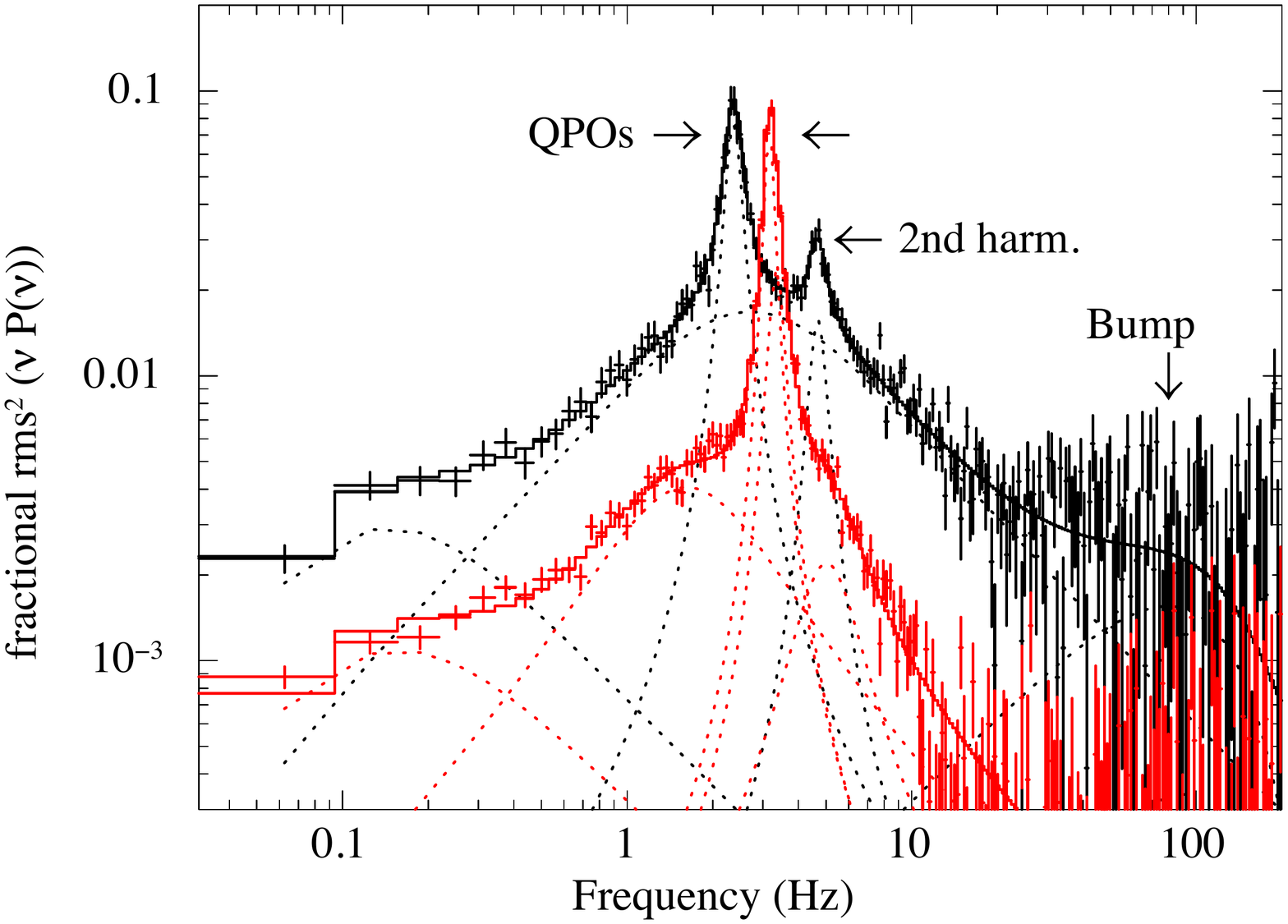}
    \includegraphics[width=0.49\textwidth ,trim= 30 30 30 30]{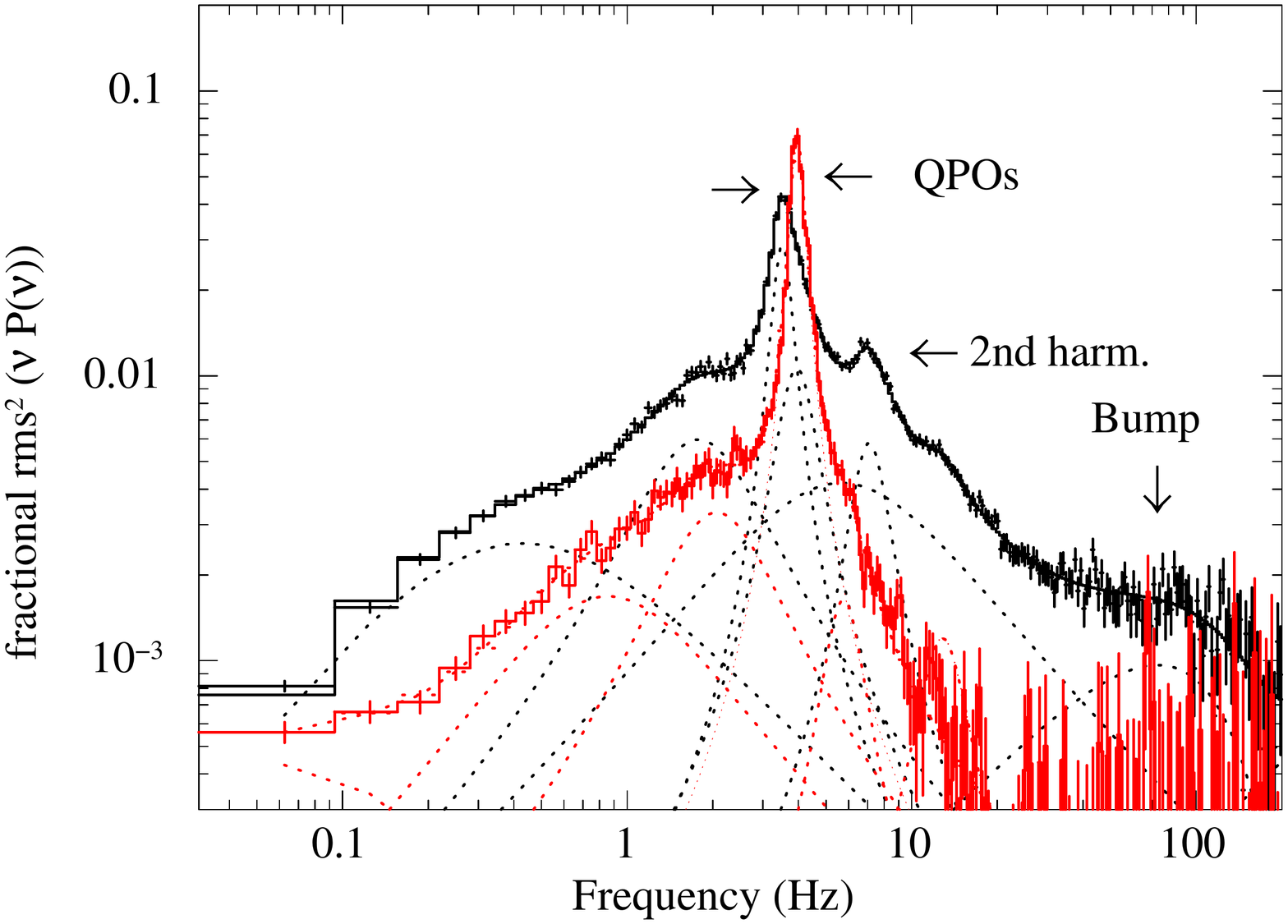}
    \includegraphics[width=0.49\textwidth ,trim= 30 30 30 30]{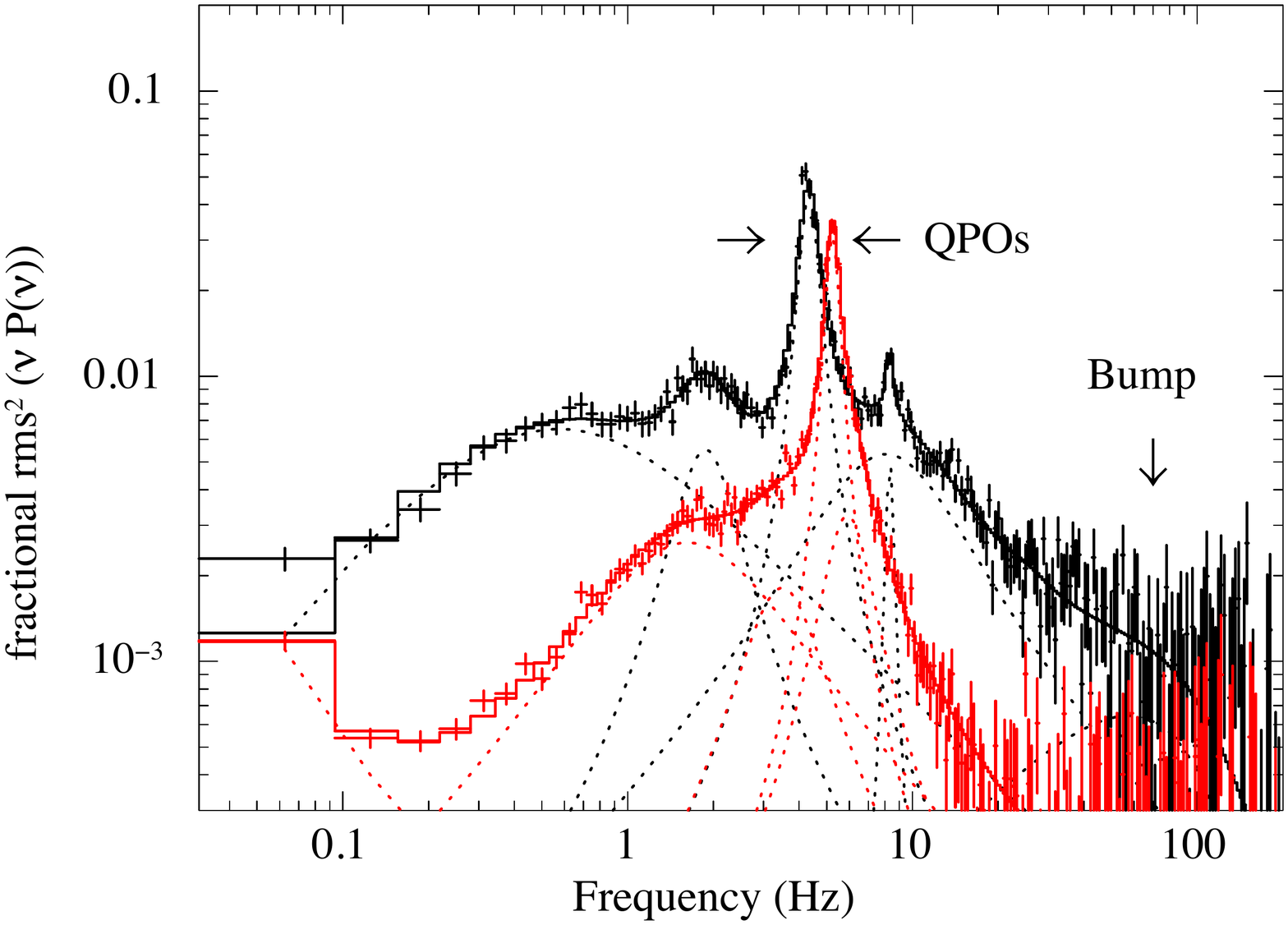}
    \caption{Representative power density spectra of observations of GRS~1915+105 with the type-C QPOs at more or less the same frequency with and without the bump. The points with error bars are the data, while the solid and dashed lines indicate the best-fitting model and the components, respectively. Top left panel: The red PDS corresponds to observation 91701-01-32-01 (QPO frequency: 1.65; HR: 0.64; rms of bump: 0.013), while the black PDS corresponds to observation 92702-01-02-01 (QPO frequency: 2.06; HR: 1.12; rms of bump: 0.074). Top right panel: The red PDS corresponds to observation 70702-01-44-00 (QPO frequency: 3.22; HR: 0.47; rms of bump: 0.012), while the black PDS corresponds to observation 92702-01-06-00 (QPO frequency: 2.36; HR: 1.11; rms of bump: 0.081). Bottom left panel: The red PDS is observation 70702-01-44-01 (QPO frequency: 3.96; HR: 0.42; rms of bump: 0.009), while the black PDS corresponds to observation 20402-01-14-00 (QPO frequency: 3.57; HR: 0.9; rms of bump: 0.063). Bottom right panel: The red PDS corresponds to observation 70702-01-45-00 (QPO frequency: 5.26; HR: 0.34; rms of bump: 0.008), while the black PDS corresponds to observation 70702-01-20-00 (QPO frequency: 4.31; HR: 0.71; rms of bump: 0.056).} 
    \label{fig:pds}
\end{figure*}

The frequency of the fundamental of the type-C QPOs of all the 410 observations is between 0.3~Hz and 6.3~Hz. Figure~\ref{fig:pds} shows selected PDS from observations with QPO frequencies being between 1.3--2.3~Hz (top left), 2.3--3.3~Hz (top right), 3.3--4.3~Hz (bottom left), and 4.3--5.3~Hz (bottom right), respectively. The top-left panel in Figure~\ref{fig:pds} shows the PDS with the bump in black, while the PDS without the bump are indicated in red. In this panel the harmonic and sub-harmonic of the type-C QPOs are clearly seen in the PDS with the bump. Note that the harmonic can also be seen in the power spectrum with the bump in the top-right and the bottom panels. The red PDS in the top-right and bottom-right panels, and the black PDS in the bottom-left panel show that two Lorentzian functions are needed to fit the QPOs, indicating that the QPOs may have shifted in frequency to some extent along the \rxte observations. Judging from these examples, the PDS with a bump possibly have more features than those without a bump. We noticed that observations with a bump have higher hardness ratio than the observations without a bump (see below), suggesting that the strength of the bump is correlated with the hardness ratio.

\subsection{Rms amplitude of the bump in individual observation}\label{subsec:hr_qpo_rms}

\begin{figure}
    \includegraphics[width=\columnwidth]{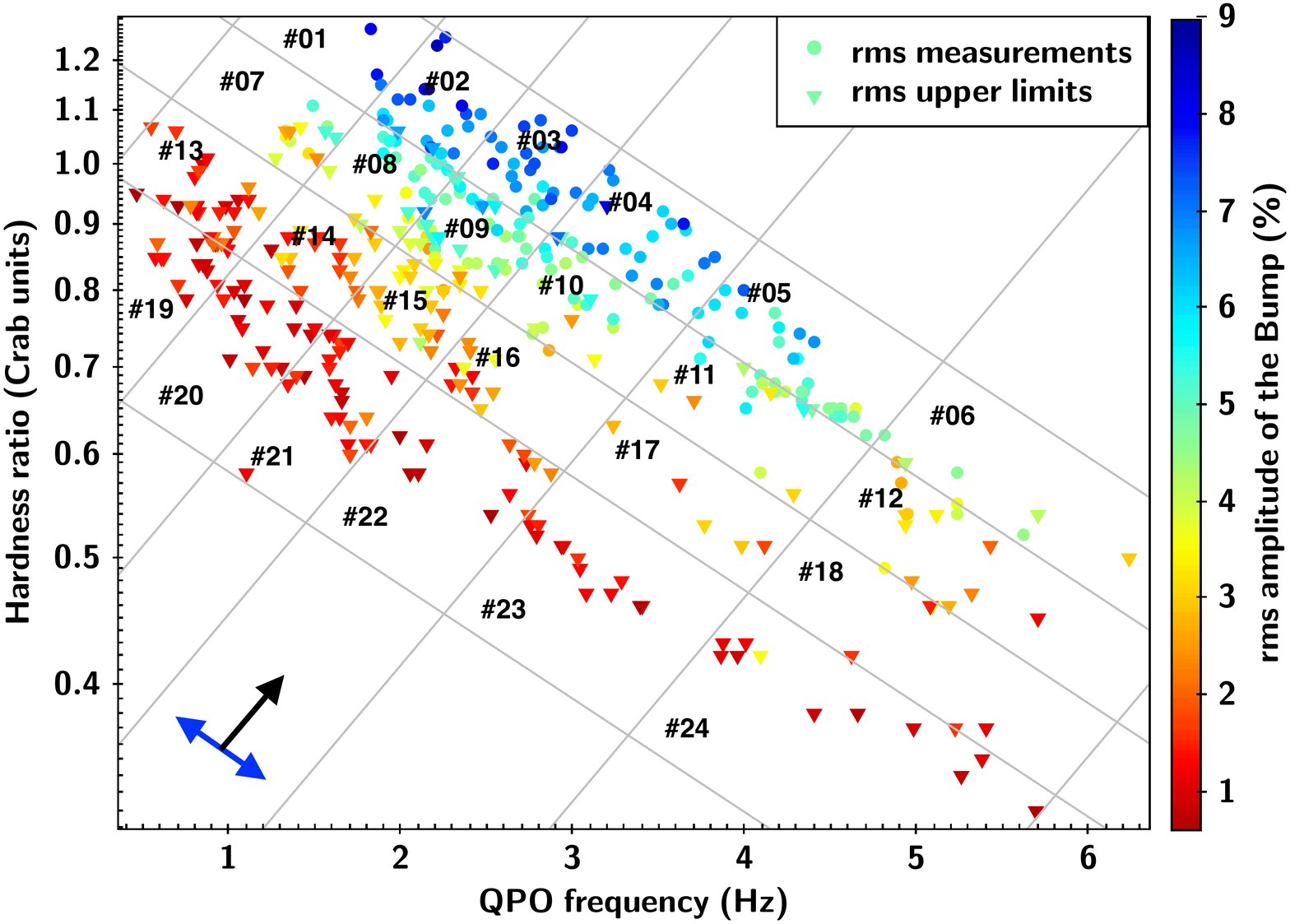}
    \caption{Hardness ratio versus QPO frequency for the 410 observations of GRS~1915+105. The colored points indicate the rms amplitude of the bump. The black and blue arrows indicate the directions in which the rms amplitude of the bump increases and remains broadly constant, respectively. The plot is divided into different regions of regular intervals on the hardness ratio and QPO frequency (see Table~\ref{tab:obs id} for details on the division and the observations in each group).}
    \label{fig:hr-qpo}
\end{figure}

\begin{figure}
    \includegraphics[width=\columnwidth , trim= 70 20 130 20]{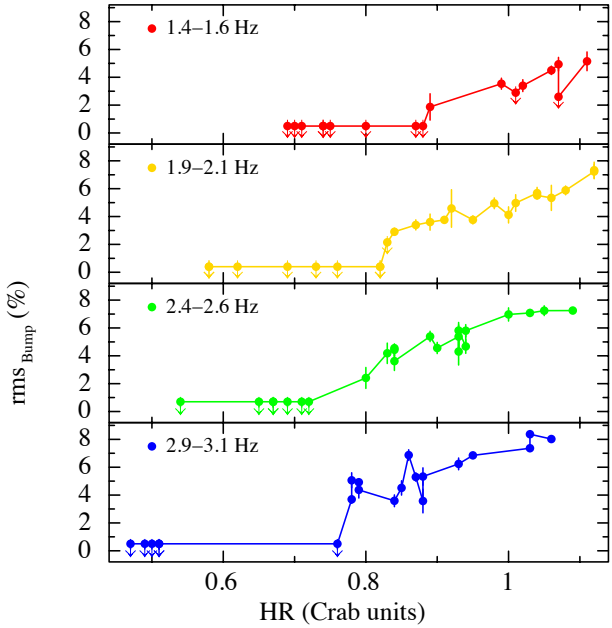}
    \caption{The rms amplitude of bump versus hardness ratio plot for selected ranges of QPO frequencies. From the top to bottom, the QPO frequencies are 1.4--1.6~Hz, 1.9--2.1~Hz, 2.4--2.6~Hz, and 2.9--3.1~Hz, respectively. The arrows indicate the 95\% upper limits.}
    \label{fig:rms-hr}
\end{figure}

\begin{figure*}
    \includegraphics[width=\columnwidth]{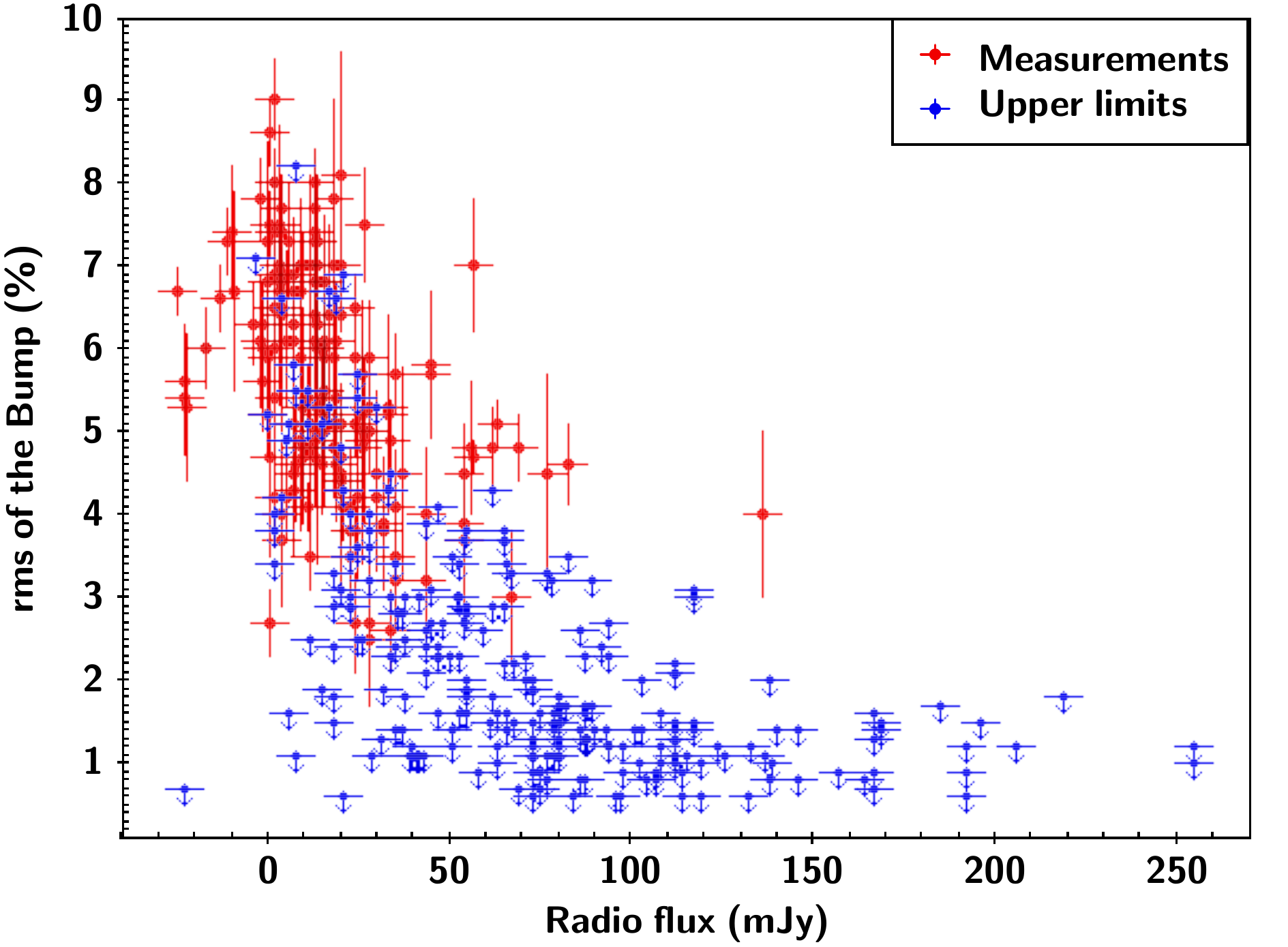}
    \includegraphics[width=\columnwidth]{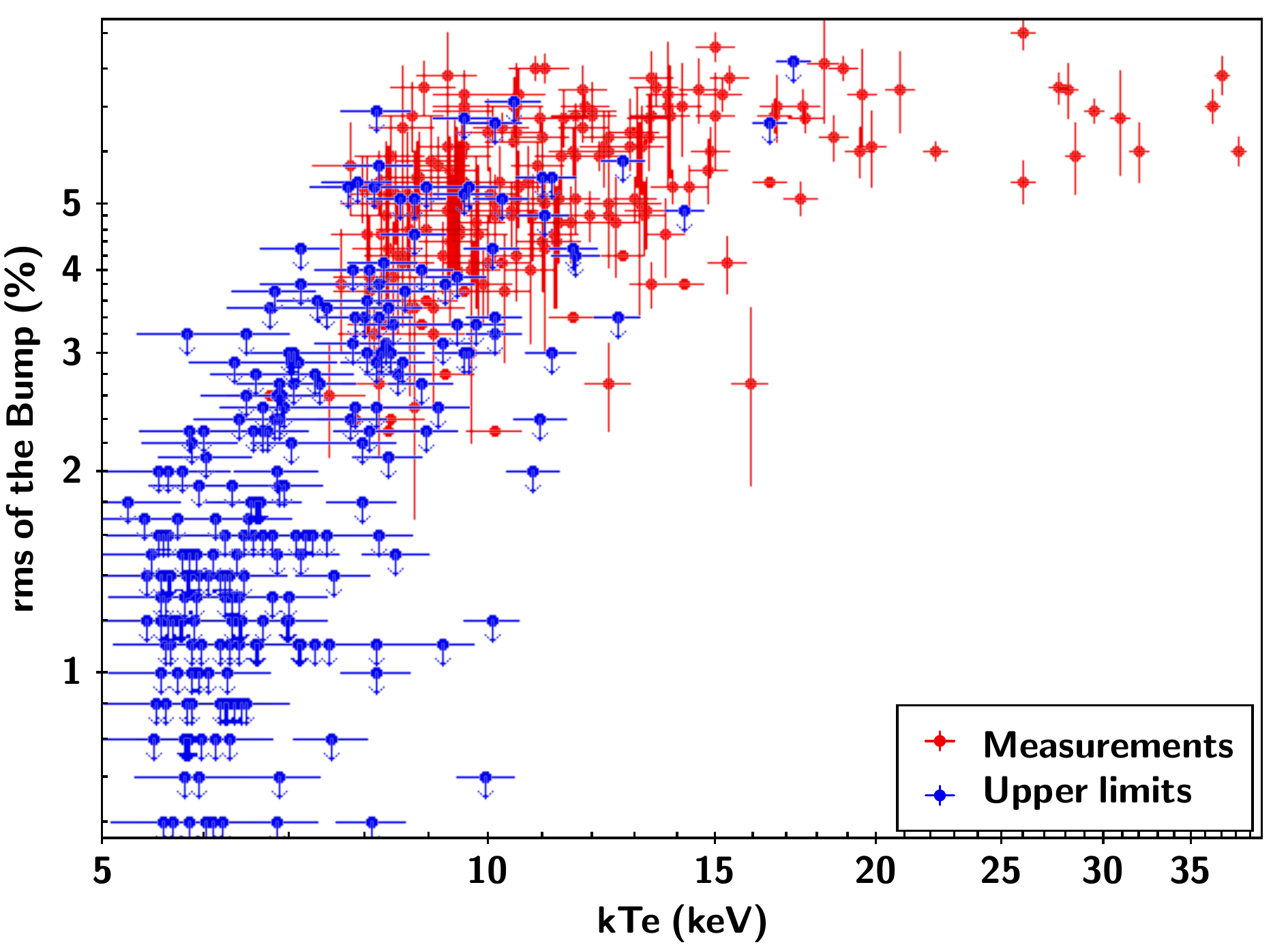}
    \caption{Left panel: The rms of the bump vs.\ the 15~GHz radio flux density for GRS~1915+105. Right panel: The rms of the bump vs.\ corona temperature, $kT_{\text{e}}$, for GRS~1915+105. In both panels, the red points show the measurements of the rms of the bump, while the blue points show the 95\% upper limit.}
    \label{fig:rms-radio}
\end{figure*}

In Figure~\ref{fig:hr-qpo} we plot the hardness ratio vs.\ the frequency of the type-C QPOs. Each point corresponds to one of the 410 \rxte observations of GRS 1915+105. Following the proposal by~\citet{2001ApJ...558..276T} that there are two types of states, corresponding to different spectral-timing states and properties in the radio band, we plot each observation in Figure~\ref{fig:hr-qpo} with the colors of the points indicating the rms amplitude of the bump. The QPO frequency varies from around 0.3~Hz to 6.3~Hz, while the hardness ratio in Crab units goes from around 0.3 to 1.3. Average errors of the QPO frequency and hardness ratio are, respectively, $\pm 0.05$~Hz and $\pm 0.001$ Crab units, both of which are smaller than the size of the data points. The rms amplitude of the bump varies in the range 1--9\%. The points plotted with circle are significant measurements of the rms amplitude of the bump, while the points marked with triangle are the 95\% upper limit of the rms amplitude since they do not fulfill the criteria in Subsection~\ref{subsec:fit}. The rms of the high-frequency bump has an average error of $\pm 0.7\%$.

Figure~\ref{fig:hr-qpo} shows a strong dependence of the rms amplitude of the bump upon hardness ratio and QPO frequency. The rms of the bump increases as both the hardness ratio and the QPO frequency increase, whereas as the QPO frequency increases and the hardness ratio decreases the rms remains broadly constant (see the black and blue arrows in Figure~\ref{fig:hr-qpo}).

In Figure~\ref{fig:rms-hr} we show the rms amplitude of the bump as a function of the source hardness ratio for four narrow ranges of the QPO frequency around, respectively, 1.5~Hz, 2~Hz, 2.5~Hz, and 3~Hz. This Figure shows that, at each QPO frequency, the rms amplitude of the bump decreases smoothly from $\sim$6--8\% down to $\sim$1--2\%  as the hardness ratio decreases. This Figure also shows that the value of the hardness ratio at which the bump becomes undetectable (the arrows in the plot are upper limits) decreases as the QPO frequency increases, consistent with the picture offered by the plot in Figure~\ref{fig:hr-qpo}. Considering the description in ~\citet{2001ApJ...558..276T} for the radio behavior, this rms evolution of the bump should show an anti-correlation with the radio flux density. The bump is strong when GRS~1915+105 is radio quiet, while the bump is weak or disappear when GRS~1915+105 is radio loud. The left panel of Figure~\ref{fig:rms-radio} shows the rms amplitude of the bump vs.\ the 15~GHz radio flux density. The rms is in the range of 1--9\%, while the radio flux density is in the range of 0--250~mJy. As the radio flux increases, the rms of the bump decreases and eventually leads to upper limits.

\subsection{Relation between the rms of the bump and the spectral parameters}

To study the relation between the bump and the spectral properties of the source, we take the parameters from the best-fitting model of a disk plus a Comptonized component, \texttt{diskbb+nthcomp}, to the 410 observations observation in GRS 1915 in~\citet{2022NatAs.tmp...51M} and~\citet{garcia}. Specifically, we plot in the right panel of Figure~\ref{fig:rms-radio} the rms amplitude of the bump vs.\ the electron temperature of the corona, $kT_{\text{e}}$ (see~\citet{2022NatAs.tmp...51M} and~\citet{garcia} for details of the spectral fitting).

From the right panel of Figure~\ref{fig:rms-radio}, it is apparent that the rms of the bump increases from $\sim$ 1\% to $\sim$ 9\% when the corona temperature $kT_{\text{e}}$ increases from 5 to 35~keV. On the contrary, we find no correlation between the rms of the bump and either $\Gamma$ or $kT_{\text{in}}$.

\subsection{Width of the bump in combined fitting}\label{width_of_the_bump}

\begin{figure*}
    \includegraphics[width=0.49\textwidth , trim=60 60 60 60]{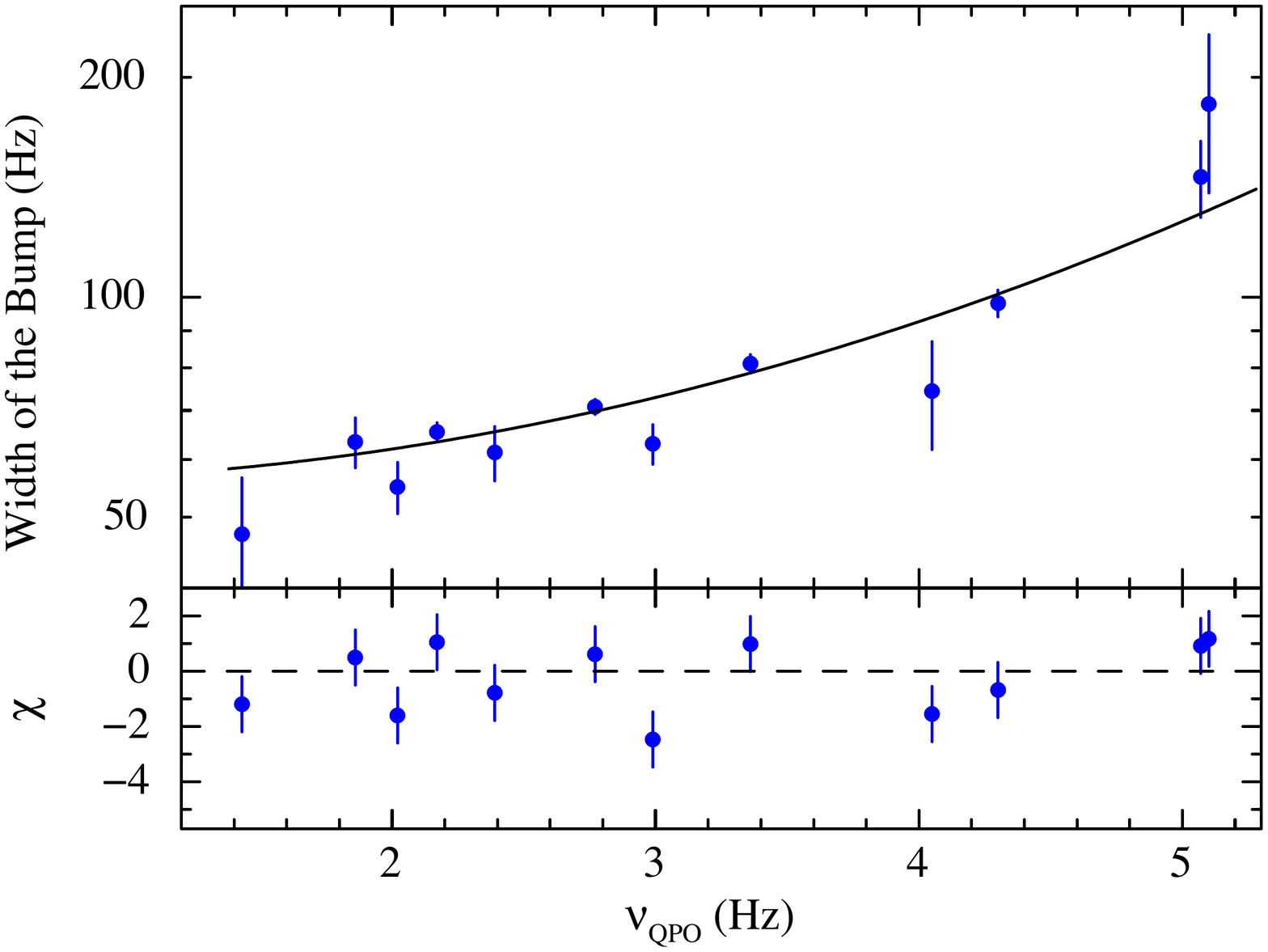}
    \includegraphics[width=0.49\textwidth , trim=60 60 60 60]{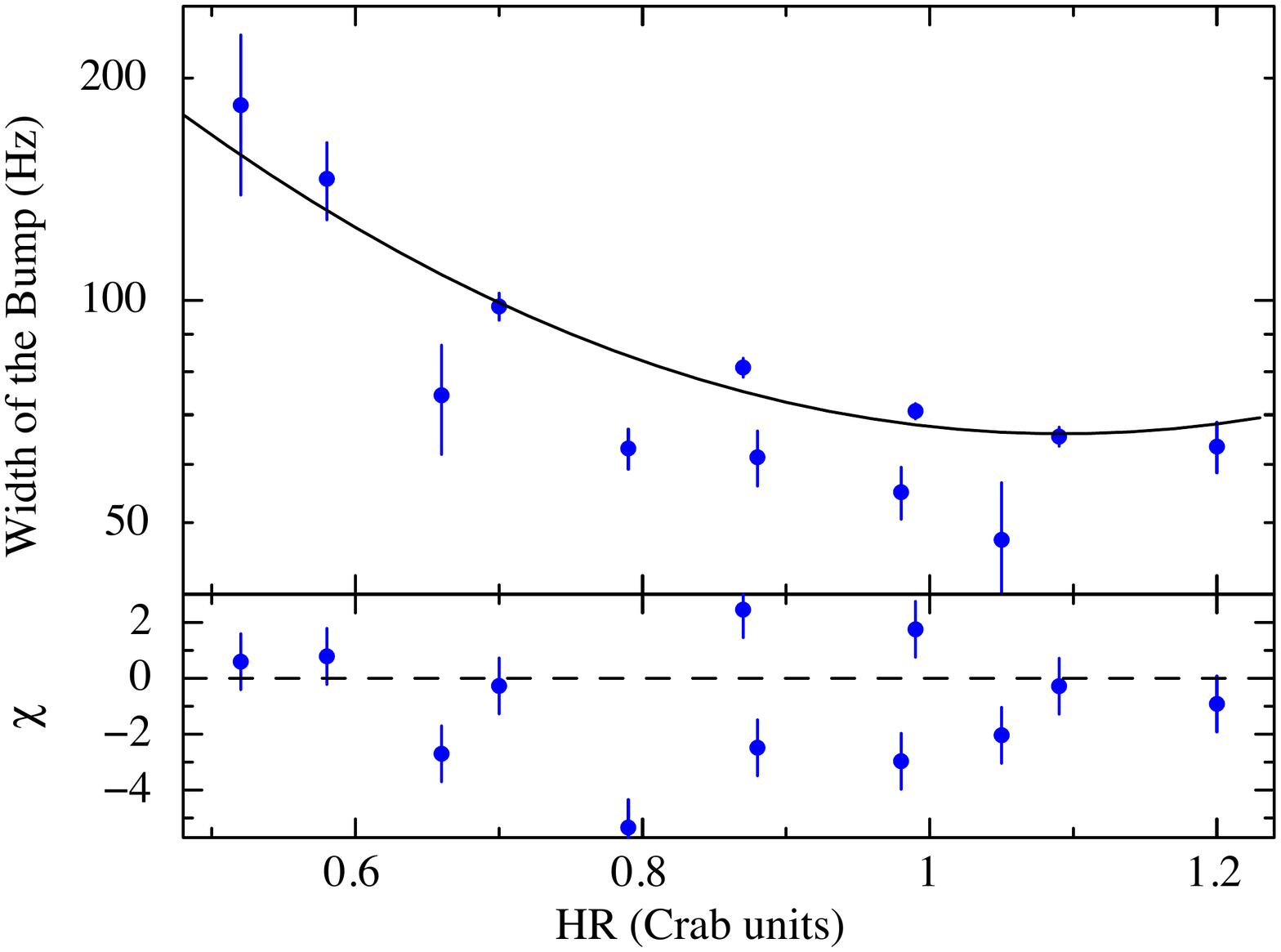}
    \caption{Width of the bump vs.\ QPO frequency (left) and hardness ratio (right) in GRS~1915+105. The error bars correspond to 90\% confidence level. The solid black line in both panels is the best-fitting quadratic function to $\log(\text{Width})$ vs. $\nu_{\text{QPO}}$ and HR, respectively. The bottom panels show the residuals given as (data $-$ model)$/$error.}
    \label{fig:evolution}
\end{figure*}

\begin{figure}
    \includegraphics[width=\columnwidth , trim=60 60 60 60]{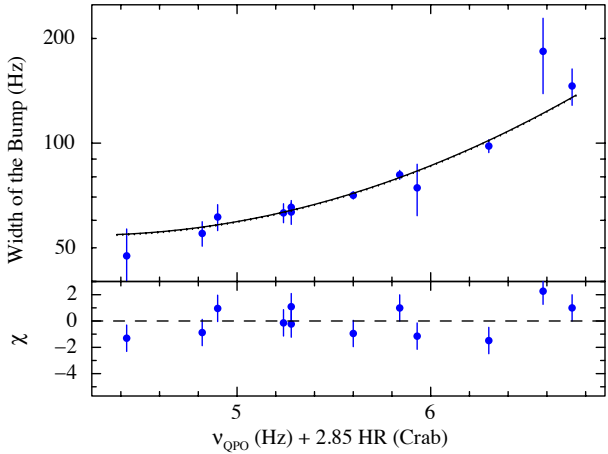}
    \caption{The width of the bump vs.\ a linear combination of the QPO frequency and hardness ratio for GRS~1915+105. The error bar corresponds to 90\% confidence level. The solid black line and the bottom panel are the same as those in Figure~\ref{fig:evolution}.}
    \label{fig:mixed_evolution}
\end{figure}

\begin{table}
    \caption{Best-fitting values of width of the bump in GRS~1915+105 vs.\ QPO frequency and HR for the groups in Figure~\ref{fig:hr-qpo}.}
    \centering
    \begin{tabularx}{\columnwidth}{llll} 
        \hline
        Group Index & QPOs Frequency (Hz) & Hardness Ratio & Width (Hz) \\
        \hline
        \#01 & 1.86 & 1.20 & $63.4_{-4.7}^{+5.3}$ \\
        \#02 & 2.17 & 1.09 & $65.4_{-1.9}^{+2.0}$ \\
        \#03 & 2.77 & 0.99 & $70.8_{-1.5}^{+1.9}$ \\
        \#04 & 3.36 & 0.87 & $81.1_{-2.3}^{+2.5}$ \\
        \#05 & 4.30 & 0.70 & $98.1_{-3.9}^{+4.4}$ \\
        \#06 & 5.07 & 0.58 & $146\pm 18$ \\
        \#07 & 1.43 & 1.05 & $47.4_{-8.4}^{+10.1}$ \\
        \#08 & 2.02 & 0.98 & $55.0_{-4.1}^{+4.8}$ \\
        \#09 & 2.39 & 0.88 & $61.3_{-4.9}^{+5.6}$ \\
        \#10 & 2.99 & 0.79 & $63.0_{-3.7}^{+4.2}$ \\
        \#11 & 4.05 & 0.67 & $74.4_{-6.5}^{+18.6}$ \\
        \#12 & 5.16 & 0.52 & $184\pm 45$ \\
        \hline
    \end{tabularx}\label{tab:sigma_qpo_hr}
\end{table}

We use a number of exponential functions to divide the hardness-QPO frequency plot into 24 regions (see Figure~\ref{fig:hr-qpo}). We combine observations and perform fitting on the combined power spectra, given that in each group of observations the properties of the bump are consistent with being the same. The division of the groups and the observation IDs in each group are listed in Table~\ref{tab:obs id}. This procedure allows us to obtain a more stringent constraint on the properties of the bump. The components we use to fit the observation in a group are the same as described in~\ref{subsec:fit}. In a square, we let all the Lorentzian functions free when we fit the power spectra, and only link all the parameters of the zero-centered Gaussian function to measure both the width and the rms of the bump for the combined observations. Since the constraint of the width can be more stringent, we relax the limit of 150~Hz which we use in the individual fittings. But if the width of the Gaussian function is unconstrained, we fix it at 70~Hz and we give an upper limit to the rms amplitude. We then use the fitting results on the full band for each data group as the baseline for the separate bands. Within one observation group, for the separate bands, we fix the position and the width of the Lorentzian functions, as well as the width of the Gaussian function, to be the same as the best-fitting values in the full band, and we fit only the rms amplitude of the components. We follow the process described in~\ref{subsec:fit} to calculate the rms as well as the 90\% error bars and 95\% confidence-level upper limits. If the calculation of the rms upper limit is larger than unity, we set the value to unity due to the fact that the rms should not be larger than unity~\citep{1990A&A...230..103B}.

In Figure~\ref{fig:evolution} we plot the width of the bump versus the QPO frequency and the hardness ratio, respectively. The width of the Gaussian function appears to depend on the QPO frequency and the hardness ratio. The width of the bump is broadly in the range of 40--230~Hz. The QPO frequency range and the hardness ratio range are 1.3--5.3~Hz and 0.5--1.3, respectively, as indicated by the group indices in Table~\ref{tab:sigma_qpo_hr}. The table shows the best-fitting value of the width of the bump in each group for the width of the bump with significant measurements, while the QPO frequency and hardness ratio correspond to the averaged values in each data group. Note that in the table we exclude the groups in which the width of the bump is fixed at 70~Hz because the rms and the width of the bump are unconstrained in those cases. In Figure~\ref{fig:evolution}, the left panel shows that the width of the bump and the QPO frequency are correlated, while the right panel shows that the width of the bump and the QPO frequency are anti-correlated. These plots suggest that there may be a parameter $c$ such that for some value of $c$ the width of the bump is more tightly correlated to a variable $x = \nu_{\text{QPO}} + c\,\text{HR}$ than to either $\nu_{\text{QPO}}$ or HR separately~\footnote{The variable $x$ represents a rotation of the $\nu_{\text{QPO}}$ and HR axes around the axis of the width of the bump, where $c=\cot{\phi}$, with $\phi$ being the rotation angle of the $\nu_{\text{QPO}}$and HR axes. The best fit yields $\phi=0.37$~rad.}. In Figure~\ref{fig:mixed_evolution} we plot the width of the bump vs.\ $x$ for the value of $c$ that gives the smallest $\chi^2$ value to the fit of a quadratic relation of the logarithm of the bump vs.\ $x$. This Figure shows that the width of the bump depends upon this specific linear combination of the QPO frequency and the hardness ratio.

\subsection{Rms spectrum of the bump}

\begin{figure*}
    \includegraphics[width=1.0\textwidth]{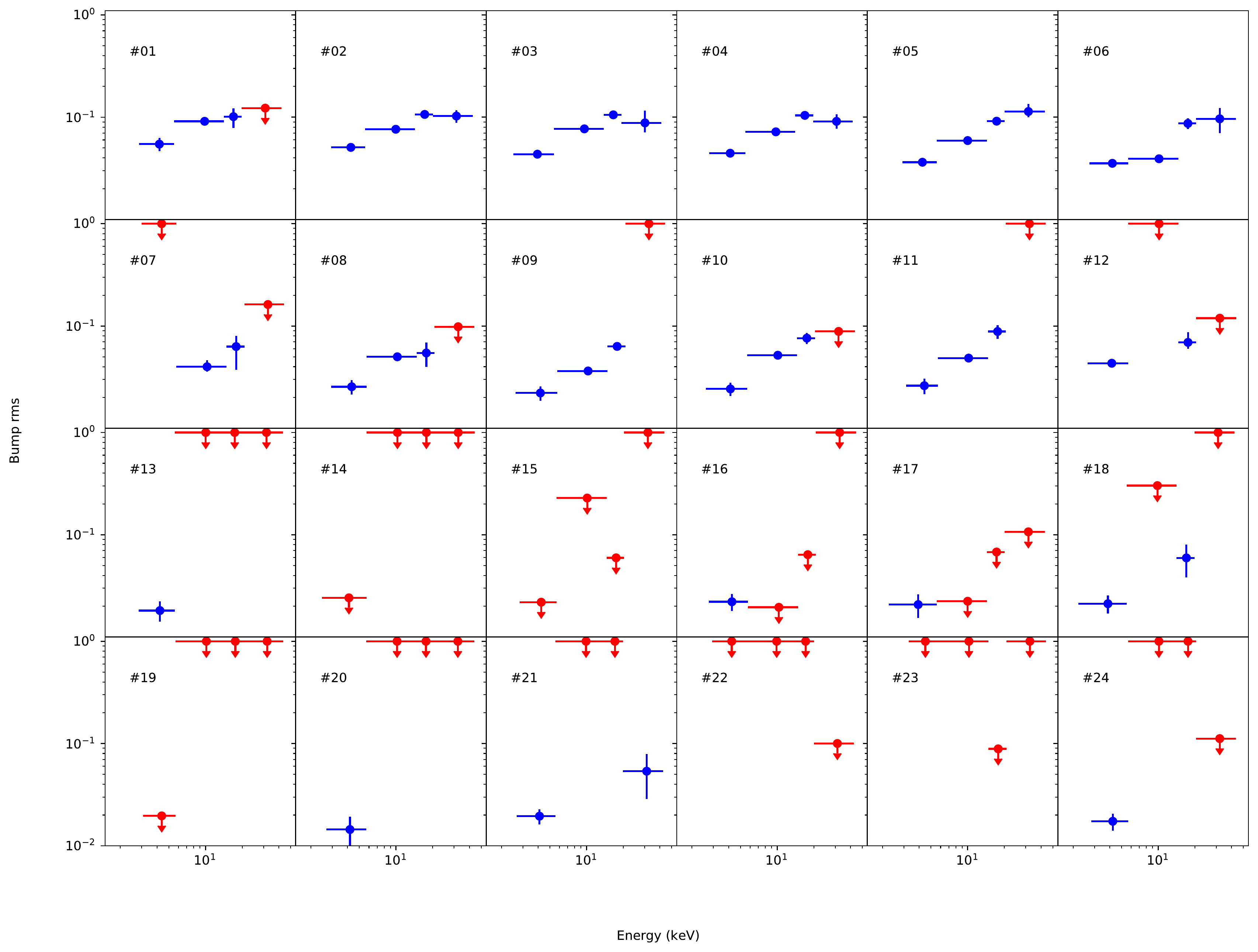}
    \caption{The rms of the bump in GRS~1915+105 versus energy. Different panels indicate the different data groups in Figure~\ref{fig:hr-qpo}, arranged in the same manner as in that Figure. The blue points are significant measurements with error bars of 90\% confidence level, while the red points are 95\% upper limits.} 
    \label{fig:rms-energy}
\end{figure*}

The rms spectra of the bump in the combined power spectra are plotted in Figure~\ref{fig:rms-energy}. For each group of combined observations, the rms of the bump generally increases with energy, from $\sim$1\% in the 3--6~keV band to $\sim$~10\% in the 20--40~keV band. In those cases in which the best-fitting values are 0 we plot the upper limits (red points) of the rms amplitude of the bump. As it is apparent in Figure~\ref{fig:rms-energy}, these broad upper limits are also consistent with this increasing trend.

In the groups \#13--\#24 (last two rows), we see that most rms values for different energy bands are upper limits. Groups \#13, \#16, \#17, \#20, and \#24 have significant measurements of the rms amplitude for the lowest energy band, where the rms amplitude is around 1\%. There are two significant measurements in groups \#18 and \#21. We find that the rms spectrum tends to have more significant data points as both the hardness ratio and QPO frequency increase. This trend is expected given the trend of the rms amplitude of the bump in Figure~\ref{fig:hr-qpo}. When we have enough measurements we see that the rms first increases with energy up tp 10--20~keV and then stays more or less constant within the errors at high energies. We also notice that if we can measure a significant rms, in the same row the rms for the same energy band remains broadly the same, while in the same column, the rms for the same energy band increases from bottom to top.

\subsection{Lags of the bump}

We calculate FFTs in each observation in two separate bands: The low band is channel 0/8 to 35, while the high band is channel 36 to 249. The coherence function over a broad high frequency range, 30--150~Hz, of this two FFTs within one observation is always consistent with zero, which means that we cannot measure the phase lags of the bump. The combined FFTs of observations in a group in Figure~\ref{fig:hr-qpo} still give a low coherence function. This implies that the phase lags of the bump at the frequency of the width cannot be measured using the current data.

\section{Discussion}\label{sec:discussion}

We have detected the high-frequency bump, with a characteristic frequency between 30~Hz and ~150Hz, in the power density spectra of GRS~1915+105 of the \rxte observations performed during the $\chi$ state when the type-C QPO is present. Not all 410 observations show the bump. The rms amplitude of the bump strongly depends on both the QPO frequency and the hardness ratio, as shown in Figure~\ref{fig:hr-qpo}. When the QPO frequency increases from 0.5 to 6.5~Hz and the hardness ratio increases from 0.3 to 1.3, the rms amplitude of the bump gradually increases from $\sim$1\% to $\sim$9\%. The radio flux of GRS~1915+105 is anti-correlated with the rms amplitude of the bump, suggesting that there is a mechanism that drives energy between the bump and the jet. The characteristic frequency of the bump shows a correlation with the frequency of the type-C QPO and an anti-correlation with the hardness ratio. The correlation is much tighter when we plot the width of the bump vs.\ a combination of the QPO frequency and the hardness ratio. The rms amplitude of the bump for each combined fitting (Figure~\ref{fig:rms-energy}) increases from around 1\% to 10\% as the photon energy increases from $\sim$3 to $\sim$30~keV.

\citet{2001ApJ...558..276T} studied the bump and classified the observations of GRS~1915+105 into two types: Type I observations correspond to the radio-quiet state showing a high frequency bump, whereas type II observations correspond to the radio loud state in which the high frequency bump is significantly weaker or absent. The presence of the bump depends upon the QPO frequency: Type I observations generally show a type-C QPO at higher frequencies than type II observations. However, after systematically exploring the datasets, we find that at a given QPO frequency, the bump can either be present or not (see Figure~\ref{fig:pds}). From Figure~\ref{fig:hr-qpo} it is apparent that the rms amplitude of the bump depends not only upon the QPO frequency but also upon the hardness ratio. The rms of the bump increases monotonically and continuously as both the QPO frequency and the hardness ratio increase, indicating that the hardness ratio and the QPO frequency both contribute to the strength of the bump. From all this, as well as Figure~\ref{fig:rms-hr} and left panel of Figure~\ref{fig:rms-radio}, it is apparent that the type-I and type-II observations are in fact the extreme cases of a continuous and smooth phenomenology, with the types I and II corresponding to the cases of the strongest bump and the strongest radio flux, respectively.

Previous studies have found a universal correlation between the X-ray and radio flux in black-hole candidates~\citep{2003MNRAS.344...60G,2013MNRAS.428.2500C,2014MNRAS.445..290G}, although GRS~1915+105 does not seem to follow this correlation~\citep{2001ApJ...556..515M}. From our result in the left panel of Figure~\ref{fig:rms-radio}, we find that in GRS~1915+105 the radio flux is anti-correlated with the rms amplitude of the bump. \citet{2022NatAs.tmp...51M} suggested that the mechanism that produces the bump in GRS 1915+105 should be connected to the corona. Specifically, Figure 1 in~\citet{2022NatAs.tmp...51M} shows that the rms of the bump increases when the corona temperature increases and the radio flux and the emission of the iron line due to reflection decrease. We show that the rms of the bump is anti-correlated with the radio flux (left panel of Figure~\ref{fig:rms-radio}) and correlated with the corona temperature (right panel of Figure~\ref{fig:rms-radio}). On the other hand, the rms of the bump is uncorrelated with the best-fitting disk temperature, indicating that the rms amplitude of the bump is not related to the disk. All this, and the fact that in those groups in Figure~\ref{fig:rms-energy} in which we can measure the rms spectrum of the bump significantly up to $\sim$ 30--40~keV, with the rms amplitude of the bump at the highest energies being $\sim$ 10\%, strongly suggest that mechanism that produces the rms amplitude of the bump is connected with corona. We note that the rms spectrum of the low-frequency QPOs in black-hole systems~\citep[e.g.,][]{2020MNRAS.494.1375Z,2020NatAs.tmp..184M} suggests that the rms spectrum of this QPO is also formed in the corona around the black hole~\citep{2020MNRAS.492.1399K,2022NatAs.tmp...51M,garcia}. Since radio emission comes from the jet which is relatively far away from the black hole, if the radio emission and the rms of the bump are physically connected, the energy of the high-frequency oscillations from the X-ray photons would have to propagate to the region where the radio emission is produced to explain this rms-radio anti-correlation. In this scenario, the decay of the rms of the bump gives rise to the radio flux. When the bump is no longer significantly detected, the radio flux increases faster, indicating that there may be another mechanism that also contributes to boosting the radio emission.

After \citet{2000MNRAS.318..361N} used multiple Lorentzian components to fit the PDS of GX~339-4, \citet{2002ApJ...572..392B} proposed a unified description of the timing features of accreting X-ray binaries. \citet{2002ApJ...572..392B} studied the high-frequency variability and extended the correlation~\citep{1999ApJ...520..262P} between low frequency QPOs ($\nu_{\text{LF}}$ in the range of around 0.1--100~Hz), low-frequency broadband ($\nu_{l}$) and high-frequency broadband ($\nu_{u}$) components. In the left panel of Figure~\ref{fig:evolution}, we confirm the correlation between $\nu_{\text{LF}}$ and $\nu_{u}$ in GRS~1915+105 with the $\nu_\text{LF}$ in the range of around 1--6~Hz, where we identify the $\nu_{u}$ component in GRS~1915+105 with the high-frequency bump. Our results show that when the QPO frequency increases, the width of the bump increases. This indicates that the QPOs and the bump may have the same origin. The properties of the spectra of the source, which in those observations are dominated by Comptonization~\citep{2001ApJ...558..276T}, also contribute to the width of the bump. The anti-correlation between the width of the bump and the hardness ratio, as shown in the right panel of Figure~\ref{fig:evolution}, indicates that the Comptonization process can somehow move the bulk of the high-frequency variability to lower frequencies, resulting in a drop of the width of the bump. The correlation between the frequency of the bump and both the QPO frequency and the hardness ratio is stronger than that that of the frequency of the bump and the QPO frequency alone. Note that in Figure~\ref{fig:hr-qpo} the rms of the bump increases when both the QPO frequency and hardness ratio increase. So considering both the effect of QPO frequency and hardness ratio, as both parameters increase, the width and the rms amplitude of the bump increase.

Since the bump is fitted with a zero-centered Gaussian, the width of the bump in GRS~1915+105 is the parameter $\sigma$ in the Gaussian function in the range of 30--250~Hz in our observations (see Figure~\ref{fig:mixed_evolution}). We regard the width of the bump as its characteristic frequency. This frequency range is remarkably similar to the frequency of the high-frequency QPOs (HFQPOs) in GRS~1915+105 at around 41~Hz and 69~Hz~\citep{1997ApJ...482..993M,2001ApJ...554L.169S,2013MNRAS.432...10B}. According to~\citet{2000A&A...355..271B}, the HFQPOs appear in the $\gamma$ class and this class reflects transition between states A and B, due to the difference in the temperature of the inner disk. This difference can contribute to the ratio between the hard and soft photons, resulting in the state transitions. The HFQPOs frequency mentioned above cannot be the Keplerian frequency at the innermost stable circular orbit (ISCO) around the black hole in GRS~1915+105~\citep[see discussion in][]{2013MNRAS.432...10B}. Instead, \citet{1998ApJ...492L..53C} suggested that the HFQPOs in GRS~1915+105 are due to the precession of the hot inner flow close to the black hole, inside the disk truncation radius. Given all this, and the rms spectrum of the HFQPOs~\citep{1997ApJ...482..993M,2013MNRAS.432...10B}, the radiative properties of the HFQPOs must arise from Comptonization in the hot electrons in the corona. This suggests that in the $\gamma$ class, where narrow HFQPOs are detected, there is a strong coupling between the disk and the corona, and when this coupling is resonant a narrow QPO appears~\citep{2013MNRAS.435.2132M}. The bump, however, appears in the $\chi$ class, which corresponds to state C~\citep{2000A&A...355..271B}. In the C state, the high-energy emission is more dominant than that in the B state~\citep{2000A&A...355..271B}, suggesting that this may be the cause of hard spectrum of the bump. In this state, the coupling between the disk and the corona would be much weaker than that in the state B, and hence the resonance between the disk and the corona would be more damped in the $\chi$ class than that in the $\gamma$ class observations, leading to a broader high-frequency variability.

HFQPOs at 66~Hz were discovered in the IX class of IGR~J17091$-$3624, which is similar to the behavior of $\gamma$ class of GRS~1915+105~\citep{2012ApJ...747L...4A}. This means that observations of IGR~J17091$-$3624 could potentially be used to test the proposed link between the bump and the HFQPOs. Testing the relation to the jet might be more challenging given that the radio flux is significantly weaker in IGR~J17091$-$3624~\citep{2011A&A...533L...4R,2020MNRAS.491.4857G}.

The rms amplitude of the bump we obtain is in the range 1 to 9\% for the significant measurements (blue points in Figure~\ref{fig:hr-qpo}), lower than the rms of the simultaneous type-C QPO (3.6--16.2\%), but nearly the same as the amplitude of the second harmonic and the subharmonic of the QPOs~\citep{2020MNRAS.494.1375Z}. This means that the bump is actually not the dominant feature in the whole power spectrum, but only dominant at high frequencies. It is worth noting that the rms amplitude of the bump increases with energy, as seen in the rms-energy spectra for the combined fittings in Figure~\ref{fig:rms-energy}. This trend is the same as that of the rms-energy spectra of the QPOs~\citep{2020MNRAS.494.1375Z}, suggesting that the QPOs and the high frequency bump may originate from the same region. Since the disk temperature in this observation is always under $\sim$2~keV~\citep{2001ApJ...558..276T} and the rms amplitude of the bump can be as large as 10\%--15\% at 30~keV, it must be the hot corona that modulates the hard photons and produces the bump. In this scenario, the soft photons interact with the hot electrons in the corona and move to high energies.

Our rms-energy spectra offer the potential to study the properties of the corona. A Comptonization model was recently proposed by~\citet{2020MNRAS.492.1399K} to fit the spectral-timing properties of QPOs~\citep[see also][]{2021MNRAS.tmp..829K}. The model provides information about the physical properties of the corona that are not directly accessible through time-averaged energy spectrum. Note that this model does not explain the dynamical origin of the QPOs, but assumes that the QPO frequency is produced by some mechanisms~\citep[e.g.,][]{2009MNRAS.397L.101I,2018ApJ...858...82Y,2020ApJ...897...27Y}. This Comptonization model was used to study the Comptonizing medium in the neutron star 4U~1636$-$53 through the lower kilohertz QPOs~\citep{2020MNRAS.492.1399K}, and the rms and lag-energy spectra of the type-B QPOs in MAXI~J1348$-$630~\citep{2021MNRAS.501.3173G} and the type-C QPOs in MAXI~J1535$-$571~\citep{2022MNRAS.512.2686Z}. Further studies are ongoing to apply this model to the LFQPOs in the black hole GRS~1915+105~\citep[see also][]{2021MNRAS.tmp..829K,garcia}. Considering that the bump may also originate from the corona through Comptonization, the model can also be applied to the high-frequency bump. The hard lags of the 67-Hz HFQPO are thought to be due to the longer path of the hard photons than the path of the soft photons~\citep{1998ApJ...492L..53C} and resonance in the corona~\citep{2013MNRAS.435.2132M}. However, it is still unclear whether the hard photons in the bump lag the soft photons because we are unable to measure the lags of the bump using the \rxte dataset. Future X-ray missions like \textit{eXTP}~\citep{2019SCPMA..6229502Z} and \textit{ATHENA}~\citep{2013arXiv1306.2307N} will be sensitive enough to measure the lags of the bump, and will help us understand the disk-corona interplay that drives the high-frequency variability in this source.

\section{Conclusions}\label{sec:conclusion}

We studied the high-frequency variability in the power density spectra of GRS~1915+105. We found that:
\begin{itemize}
\item The rms amplitude of a high-frequency bump with frequencies between $\sim$30 and $\sim$150~Hz in the X-ray power density spectra depends both on the frequency of the type-C QPO and the hardness ratio of the source.  
\item The rms amplitude of the bump is correlated with the temperature of the corona and anti-correlated with the simultaneous 15-GHz radio flux density of the source. 
\item The rms spectrum of the bump is hard, with the rms amplitude increasing from $\sim$1--2\% at $\sim$3~keV up to $\sim$10--15\% at $\sim$30~keV.
\item The characteristic frequency of the bump correlates much better with a combination of the frequency of the type-C QPO and the hardness ratio of the source than with the QPO frequency alone.
\item Due to the low intrinsic coherence of the variability at the frequency of the bump in the cross spectrum we could not measure the phase lags of the bump.
\end{itemize} 
We suggest that the high-frequency bump in the $\chi$ class and the 67-Hz HFQPO in the $\gamma$ class are the same variability component that is either broad or narrow depending upon the properties of the corona in these two classes. The anti-correlation between the rms amplitude of the bump and the radio emission suggests that in GRS 1915+105 the radio jet and and X-ray corona are physically connected.


\section*{Acknowledgements}

Part of this work was done by YZ during a visit at Fudan University in 2020. YZ acknowledges support from China Scholarship Council (CSC 201906100030) and the hospitality of professor Cosimo Bambi. MM acknowledges the research programme Athena with project number 184.034.002, which is (partly) financed by the Dutch Research Council (NWO). FG acknowledges support by PIP 0113 (CONICET). FG is a CONICET researcher. This work received financial support from PICT-2017-2865 (ANPCyT). LZ acknowledges support from the Royal Society Newton Funds. DA acknowledges support from the Royal Society. TMB acknowledges financial contribution from the agreement ASI-INAF n.2017-14-H.0, PRIN-INAF 2019 N.15, and thanks the Team Meeting at the International Space Science Institute (Bern) for fruitful discussions.


\section*{Data Availability}

The X-ray data used in this article are accessible at NASA's High Energy Astrophysics Science Archive Research Center \url{https://heasarc.gsfc.nasa.gov/}. The radio data used in this article are available through the website \url{http://www.mrao.cam.ac.uk/~guy/} or \url{http://www.astro.rug.nl/~mariano/GRS_1915+105_Ryle_data_1995-2006.txt}.




\bibliographystyle{mnras}
\bibliography{reference} 



\appendix

\section{Individual Observations and Group Information}

\begin{table*}
    \caption{Information about the groups defined in Figure~\ref{fig:hr-qpo}}\label{tab:obs id}
    \centering
    \begin{threeparttable}
        \begin{tabularx}{\textwidth}{lX} 
        \hline
        Group & Observation IDs\\
        \hline
        \#01 & 30703-01-34-00, 70702-01-11-00, 70702-01-11-01\\
        \#02 & 20402-01-15-00, 20402-01-19-00, 50405-01-02-01, 50405-01-02-00, 92702-01-02-01, 92702-01-02-00, 92702-01-06-00, 30703-01-41-00, 50703-01-66-00, 50405-01-02-02, 30703-01-35-00, 50703-01-66-01, 50703-01-65-01, 30703-01-24-00, 70702-01-09-00, 40703-01-01-00, 50703-01-59-01, 60405-01-04-00, 50703-01-59-00, 70702-01-10-00, 60405-01-04-03, 50703-01-64-00\\
        \#03 & 20402-01-12-00, 20402-01-24-00, 20402-01-16-00, 30402-01-16-00, 40403-01-01-00, 20402-01-11-00, 20402-01-10-00, 20402-01-09-00, 30402-01-18-00, 92702-01-05-00, 30703-01-24-02, 30703-01-24-01, 90701-01-26-00, 92702-01-03-00, 70702-01-12-01, 20402-01-05-00, 50405-01-03-00, 30703-01-36-00, 70702-01-14-00, 40703-01-09-00, 70702-01-12-00, 60405-01-03-00, 60405-01-04-05\\
        \#04 & 20402-01-20-00, 20402-01-18-00, 20402-01-07-00, 70702-01-18-01, 70702-01-18-00, 20402-01-21-00, 20402-01-21-01, 20402-01-14-00, 20402-01-13-00, 92702-01-01-01, 92702-01-01-00, 20402-01-08-01, 60405-01-04-06, 70702-01-21-01, 50703-01-25-01, 30703-01-33-00, 92702-01-09-00, 80701-01-12-00, 50703-01-08-00, 92702-01-08-03, 50703-01-28-02, 30402-01-17-00, 70702-01-21-00, 40703-01-02-00, 70702-01-23-00, 90701-01-18-00, 50703-01-28-01\\
        \#05 & 92702-01-08-00, 20402-01-08-00, 30703-01-31-00, 92082-03-01-00, 90105-02-04-00, 50703-01-28-00, 20402-01-27-01, 92702-01-08-02, 92702-01-08-01, 40703-01-05-00, 70702-01-20-01, 70702-01-20-00, 60100-01-02-00, 60702-01-01-02, 40403-01-11-00, 50703-01-62-03, 60702-01-01-00, 50703-01-62-01, 50703-01-62-00, 20402-01-04-00, 60701-01-06-01, 60701-01-04-00, 50703-01-62-02, 70702-01-22-00\\
        \#06 & 60702-01-01-10, 40403-01-10-00, 10408-01-42-00, 50703-01-48-00, 50703-01-40-01$^{*}$, 40703-01-20-01, 50703-01-41-00$^{*}$\\
        \#07 & 70702-01-07-00, 70702-01-06-01, 50703-01-50-00, 70702-01-06-00, 50703-01-50-01, 50703-01-43-01, 50703-01-43-00, 50703-01-50-02, 50703-01-43-02, 40116-01-01-05, 40116-01-01-06, 91701-01-49-00, 40116-01-01-04\\
        \#08 & 50703-01-54-01, 50703-01-54-02, 50703-01-54-00, 50703-01-52-02, 50703-01-67-02, 50703-01-67-00, 50703-01-52-01, 60100-01-01-00, 50703-01-67-01, 50703-01-52-00, 90701-01-19-00, 60100-01-01-01, 40116-01-01-07, 60701-01-01-01, 60405-01-01-01, 50703-01-51-01, 50703-01-51-00, 30182-01-01-01, 40703-01-50-02, 40703-01-16-03, 40403-01-09-00, 40116-01-01-00\\
        \#09 & 60405-01-04-04, 60405-01-04-02, 60100-01-03-01, 50703-01-53-00, 60701-01-01-00, 60405-01-04-08, 60405-01-04-01, 91701-01-22-00, 50703-01-25-02, 50703-01-25-00, 40703-01-50-00, 91701-01-50-00, 60405-01-04-07, 50703-01-53-01, 40703-01-50-01, 50703-01-53-02, 30703-01-23-01, 50703-01-55-00, 40703-01-51-03, 40703-01-51-01, 40703-01-49-02, 40703-01-49-01, 30182-01-01-00, 40116-01-02-01, 40116-01-01-03, 80701-01-11-01, 40703-01-21-00, 40116-01-02-03, 40116-01-02-02, 40703-01-15-01, 70702-01-24-00, 50703-01-46-00, 40703-01-49-00, 40703-01-21-02, 40703-01-21-01, 40116-01-01-02, 50703-01-49-00, 50703-01-44-01, 40116-01-02-00, 40703-01-15-02, 40116-01-01-01, 30182-01-02-01\\
        \#10 & 50703-01-55-01, 80701-01-11-00, 40703-01-51-02, 80701-01-11-02, 60405-01-01-00, 40703-01-51-00, 60701-01-05-01, 50703-01-56-00, 60701-01-05-00, 30703-01-23-00, 50703-01-46-01, 30182-01-04-02, 40116-01-02-07, 40116-01-02-06, 10408-01-45-00, 40703-01-17-00, 40116-01-02-05\\
        \#11 & 50703-01-27-03, 50703-01-46-02, 60702-01-01-01, 50703-01-24-01, 60701-01-06-02, 60701-01-06-00, 40116-01-02-04, 91701-01-30-01, 91701-01-30-00, 70702-01-25-00, 20402-01-02-01, 50703-01-45-00, 70702-01-25-01, 60701-01-04-02, 50703-01-45-01, 50703-01-41-03, 50703-01-41-02\\
        \#12 & 20402-01-01-00, 50703-01-40-02$^*$, 70702-01-04-00$^*$, 50703-01-41-01$^*$, 50703-01-40-03, 40703-01-20-00$^*$, 70702-01-04-01, 70702-01-04-02$^*$, 90105-01-03-01, 90105-02-03-00, 90105-02-01-00$^*$, 90105-02-01-01$^*$\\
        \#13 & 90701-01-03-00, 90701-01-05-00, 30703-01-20-00, 70702-01-01-00, 20402-01-50-00, 90701-01-04-00, 80188-03-01-01, 90108-01-06-00, 80701-01-01-00, 60701-01-29-00, 20402-01-50-01, 90701-01-02-03, 90701-01-02-00, 80188-02-01-00, 60701-01-33-01, 10258-01-02-00, 90701-01-02-02, 90701-01-02-01, 80701-01-01-01, 70702-01-08-00, 60701-01-33-00, 70702-01-08-01, 90701-01-01-00, 80188-01-01-03, 80188-01-01-04, 80188-01-01-00\\
        \#14 & 40703-01-16-02, 90108-01-04-00, 40703-01-16-01, 80701-01-01-02, 40703-01-44-03, 20402-01-51-00, 90108-01-05-00, 40703-01-45-00, 40703-01-16-00, 30703-01-19-00, 91701-01-44-00, 40703-01-45-01, 40703-01-42-03, 40703-01-42-01, 80701-01-02-00, 40703-01-42-02, 91701-01-45-00, 90108-01-03-00, 40703-01-42-00\\
        \#15 & 40703-01-47-00, 50703-01-42-01, 50703-01-42-02, 50703-01-42-00, 91701-01-44-01, 91701-01-14-01, 70702-01-05-03, 30703-01-22-01, 30703-01-22-00, 30703-01-21-00, 80701-01-02-01, 70702-01-05-01, 70702-01-05-02, 40703-01-41-05, 80701-01-02-02, 90701-01-07-01, 90701-01-07-00, 40703-01-43-00, 40703-01-41-04, 91701-01-43-01\\
        \#16 & 70702-01-03-01, 70702-01-03-00, 40703-01-17-01, 70702-01-05-00, 90701-01-06-00, 90108-01-02-00, 90412-01-01-00, 40703-01-44-02, 40703-01-44-01, 91701-01-43-00, 91701-01-42-01, 40703-01-44-00\\
        \#17 & 30703-01-18-00, 90701-01-08-00, 80127-05-05-01, 80127-05-05-02, 80127-05-05-00\\
        \#18 & 90105-01-03-00, 90105-02-02-00, 90105-02-01-02, 20186-03-02-05, 60701-01-31-01, 20402-01-47-01\\
        \#19 & 60701-01-28-00, 90701-01-01-02, 80188-01-01-01, 70702-01-51-00, 60701-01-24-02, 60701-01-24-01, 30703-01-17-00, 80188-01-01-02, 60701-01-25-01, 60701-01-25-00, 60701-01-24-00, 70702-01-50-01, 70702-01-50-00, 60701-01-26-00, 60701-01-26-01, 91701-01-03-01, 60701-01-26-02, 91701-01-03-00\\
        \#20 & 70702-01-54-01, 80127-02-02-01, 80127-02-03-00, 60701-01-26-03, 70702-01-54-00, 70702-01-52-00, 80127-02-01-01, 80127-02-02-02, 40703-01-41-01, 30703-01-16-00, 40703-01-41-00, 91701-01-04-00, 70702-01-47-01, 91701-01-02-02, 91701-01-02-00, 90105-05-03-05, 70702-01-47-00\\
        \#21 & 40703-01-41-02, 30402-01-12-02, 30402-01-12-03, 30402-01-12-01, 40703-01-41-03, 60701-01-30-01, 60701-01-30-02, 90105-05-03-04, 30703-01-15-00, 30402-01-12-00, 91701-01-02-01, 60701-01-30-00, 30703-01-14-00, 30402-01-10-00, 91701-01-32-01, 91701-01-32-00, 91701-01-07-00, 70702-01-48-01, 90701-01-50-00, 90701-01-48-00, 70702-01-48-00, 90701-01-50-01, 80127-02-02-00\\
        \#22 & 30402-01-09-00, 70702-01-32-02, 30402-01-09-01, 70702-01-32-01, 91701-01-42-00, 20402-01-49-01, 70702-01-32-00, 60701-01-23-01, 60701-01-23-00, 70703-01-01-14, 70703-01-01-08\\
        \#23 & 70703-01-01-10, 70703-01-01-13, 70703-01-01-12, 91701-01-34-00, 91701-01-34-01, 70703-01-01-07, 70703-01-01-11, 70703-01-01-05, 91701-01-06-00, 91701-01-31-00, 70702-01-44-00, 80701-01-40-01, 70703-01-01-04\\
        \#24 & 90105-07-02-00, 90105-07-01-00, 90105-07-03-00, 70703-01-01-03, 70702-01-44-01, 70702-01-46-00, 20402-01-48-00, 70703-01-01-02, 60701-01-20-00, 70702-01-45-00, 91701-01-33-01\\
        \hline
        \end{tabularx}
        \begin{tablenotes}
            \item[1] The observation IDs marked with $^*$ in groups \#01--\#06 are the insignificant measurements for the bump. Thus these observations are excluded in the combined fitting.
        \end{tablenotes}
    \end{threeparttable}
\end{table*}


\bsp	
\label{lastpage}
\end{document}